\documentclass[journal]{IEEEtran}
\usepackage{cite}

\ifCLASSINFOpdf
  \usepackage[pdftex]{graphicx}
\else
  \usepackage[dvips]{graphicx}
  \DeclareGraphicsExtensions{.eps}
\fi
\usepackage{url}

\usepackage[cmex10]{amsmath}
\usepackage[tight,footnotesize]{subfigure}

\usepackage{fixltx2e}

\usepackage{stfloats}
\begin{document}
%
\title{Comparison of ultra-stable radio frequency signals synthesized from independent secondary microwave frequency standards}
%
%
%

\author{John~G.~Hartnett, Stephen~R.~Parker, Eugene~N.~Ivanov, Travis~Povey, Nitin~R.~Nand and Jean-Michel~le~Floch 
\thanks{John~G.~Hartnett, Stephen~R.~Parker, Eugene~N.~Ivanov, Travis~Povey, Nitin~R.~Nand and Jean-Michel~le~Floch are with the School of Physics, the University of Western Australia, Crawley, 6009, W.A., Australia. J.G.H also holds a position with the Institute
of Photonics and Advanced Sensing (IPAS) and the School of Chemistry and
Physics at the University of Adelaide, Australia. }
\thanks{Manuscript received February 1, 2013; ....}}

%
%

\markboth{IEEE Trans. on Ultrasonics, Ferroelectric and Frequency Control,~Vol.~XX, No.~X, June~2013}%
{Shell \MakeLowercase{\textit{et al.}}: Bare Demo of IEEEtran.cls for Journals}
%



\maketitle

\begin{abstract}
The phase noise and frequency stability measurements of 1 GHz, 100 MHz, and 10 MHz signals are presented which have been synthesized from microwave cryogenic sapphire oscillators using ultra-low-vibration pulse-tube cryocooler technology. We present the measured data using independent cryogenic oscillators for the 100 MHz and 10 MHz synthesized signals, whereas previously we only estimated the expected results based on residual phase noise measurements, when only one cryogenic oscillator was available. In addition we present the design of a 1 GHz synthesizer using a Crystek voltage controlled oscillator phase locked to 1 GHz output derived from a cryogenic sapphire oscillator. 
\end{abstract}


%
\IEEEpeerreviewmaketitle

\begin{figure}[!t]
\centering
\includegraphics[width=3.5in]{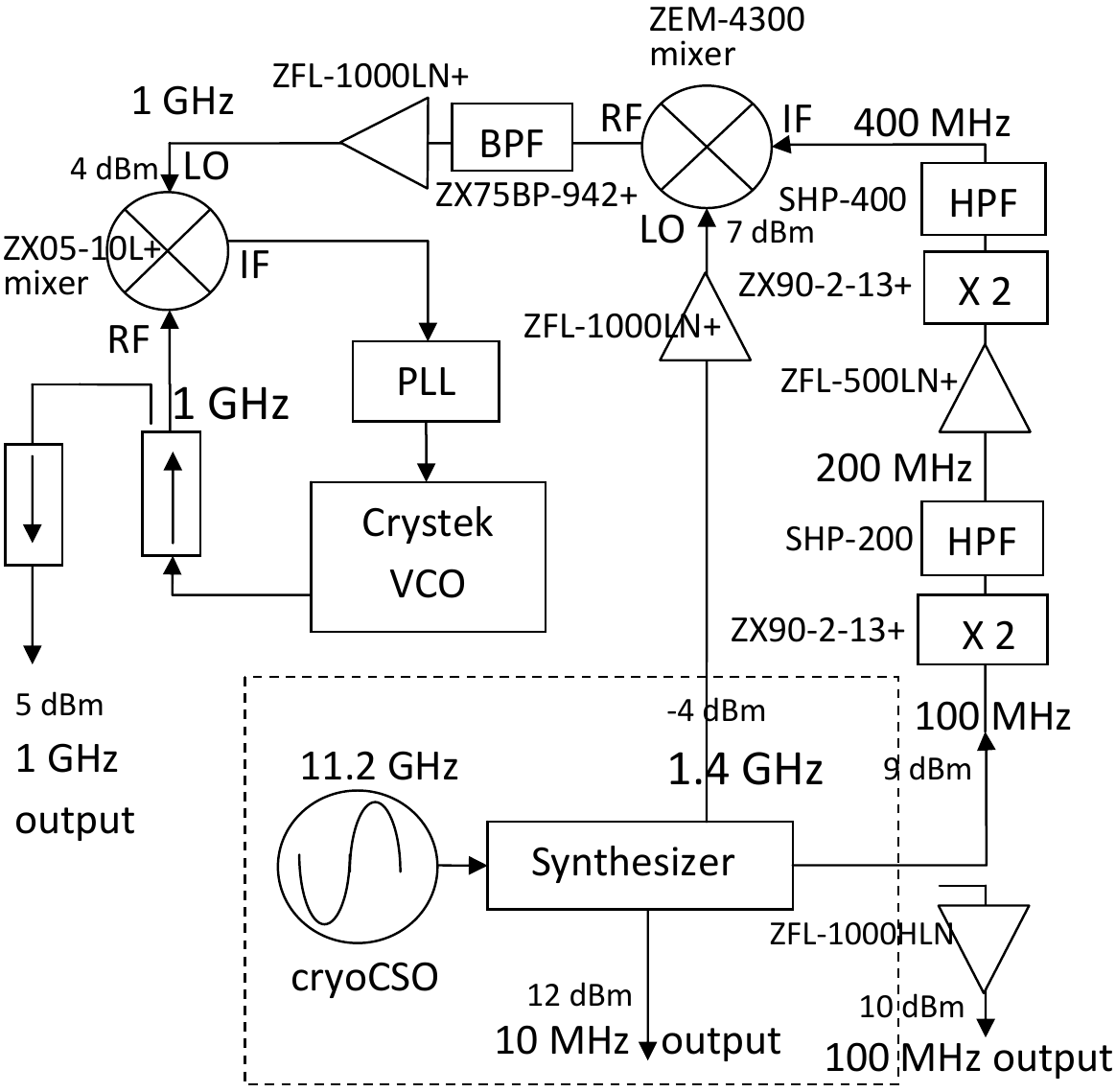}
\caption{Block diagram showing 1 GHz synthesizer with inputs of 1.4 GHz and 100 MHz from the synthesizer developed in Nand et al.~\cite{Nand2011}. The synthesizer incorporates a Crystek VCO with custom made PLL control circuit. The components in the dashed rectangle represent a cryocooled sapphire oscillator and the previously reported synthesizer~\cite{Nand2011} with 1.4 GHz, 100 MHz and 10 MHz outputs. Attenuators are not shown. HPF = High Pass Filter. x2 indicates frequency multiplier. MiniCircuits part numbers are indicated. }
\label{fig1}
\end{figure}

\section{Introduction}
\IEEEPARstart{W}{ith} the development of ultra-stable cryogenic sapphire oscillators,  using an ultra-low-vibration cryocooler and a custom designed cryostat~\cite{Wang2010} delivering unprecedented performance~\cite{Hartnett2010a, Hartnett2010b, Hartnett2012} in a robust low maintenance package, has come the need to produce user frequencies both in the RF and millimeter-wave range from the X-band signal with little or no degradation to the signal quality. Techniques to produce very low phase noise signals have been reported~\cite{Watabe2008,Hati2012,Nand2011,chambon1,chambon2,boudot}. 

The motivation for this in part has come from the very-long-baseline-interferometry (VLBI) radio-astronomy community~\cite{Doeleman2011} where it has been shown that one may make as much as a 100\% or more improvement on image quality by increasing the target angular resolution at frequencies of 350 GHz and above as compared to when a hydrogen maser is used as the frequency reference~\cite{Rioja2012}. 

The short-term fractional frequency stability of the cryocooled sapphire oscillator is as much as 100 times better than a hydrogen maser. And it has been shown that signals generated at 100 MHz from a cryocooled sappire oscillator have the potential to replace the hydrogen maser as the reference of choice. The signal degradation at 100 MHz is estimated to be much less than that at 10 MHz due to the intrinsic noise of the components used~\cite{Nand2011}. 

The X-band signal from the cryocooled sapphire oscillator is not easily distributed over long distances but as previously shown shown~\cite{Nand2011} significant loss of performance occurs when much lower frequencies are synthesized.  To synthesize 1 GHz signals with fractional frequency stability of only 1 part in $10^{15}$ could benefit both the frequency metrology community~\cite{Watabe2008} and the VLBI radio-astronomy community. This may mean significant performance is retained while 1 GHz can be distributed around a user facility reasonably well. In the case of the application to VLBI radio-astronomy it would only be a benefit if it could be shown that better stability was needed than that of the synthesized 100 MHz signal and that new methods to further suppress the instabilities arising from phase decoherence effects of the atmosphere are developed~\cite{Rioja2012}. 

In this paper we report on the design, development and evaluation of a 1 GHz synthesizer where a voltage controlled oscillator (VCO) is phase locked to a 1 GHz signal synthesized from a cryocooled sapphire oscillator.  In addition we report on the evaluation of 1 GHz, 100 MHz and 10 MHz signals synthesized from the same cryocooled sapphire oscillator and compare their phase noise and frequency stability, expanding on the work reported in Nand et al.~\cite{Nand2011}. In \cite{Nand2011} we did not have a second independent cryocooled sapphire oscillator and only reported on the residual phase noise of the 100 MHz and 10 MHz synthesizers, with an estimate of their expected performance.

Previously, a method was demonstrated where by taking the beat note between two oscillators and engineering a relatively low noise reference through frequency division of a signal derived from one of the oscillators, one can use a Symmetricom 5125A signal test set to measure the phase noise and frequency stability of oscillators operating at X-band frequencies~\cite{Hartnett2012}.  This method is limited to the case where the beat note between the signals from the two oscillators being compared falls within the measurement bandwidth of the test set being used. In the case of the two signals at precisely 1 GHz with little tunability this method was not suitable so we developed a new technique, also using the Symmetricom 5125A test set, that allows one to evaluate both the phase noise and frequency stability of the 1 GHz signals   where nominally identical synthesis chains are pumped by an X-band signal from nominally identical and independent cryocooled sapphire oscillators.

\section{Methods} 
 
In \cite{Nand2011} we employed low-phase-noise digital frequency dividers to produce a low-phase-noise signal by phase locking the signal from a low-phase-noise 100 MHz quartz oscillator to that of the cryogenic sapphire oscillator. By introducing a direct digital synthesizer (DDS) unit mixed into the microwave signal before frequency division we were able to achieve 14 significant figures of frequency tuning. 

However we were only able to report the residual phase noise of the synthesizer due to the fact that at that time we did not have a second operational cryocooled sapphire oscillator. In this paper we report on the measurement of these synthesized signals using two nominally identical systems -- two independent cryogenic sapphire oscillators using an ultra-low vibration custom designed cryostat (cryoCSOs) -- each providing the input signal near 11.2 GHz to its own frequency synthesizer.

From the 1.4 GHz and 100 MHz outputs of the synthesizer reported in \cite{Nand2011} a 1 GHz output was synthesized. This design is shown in Fig. \ref{fig1}. To this was added a custom designed phase-lock loop (PLL) circuit and a low-phase noise Crystek 1 GHz VCO (model no. CRBV55CX-1000-1000).

\begin{figure}[!t]
\centering
\includegraphics[width=3.5in]{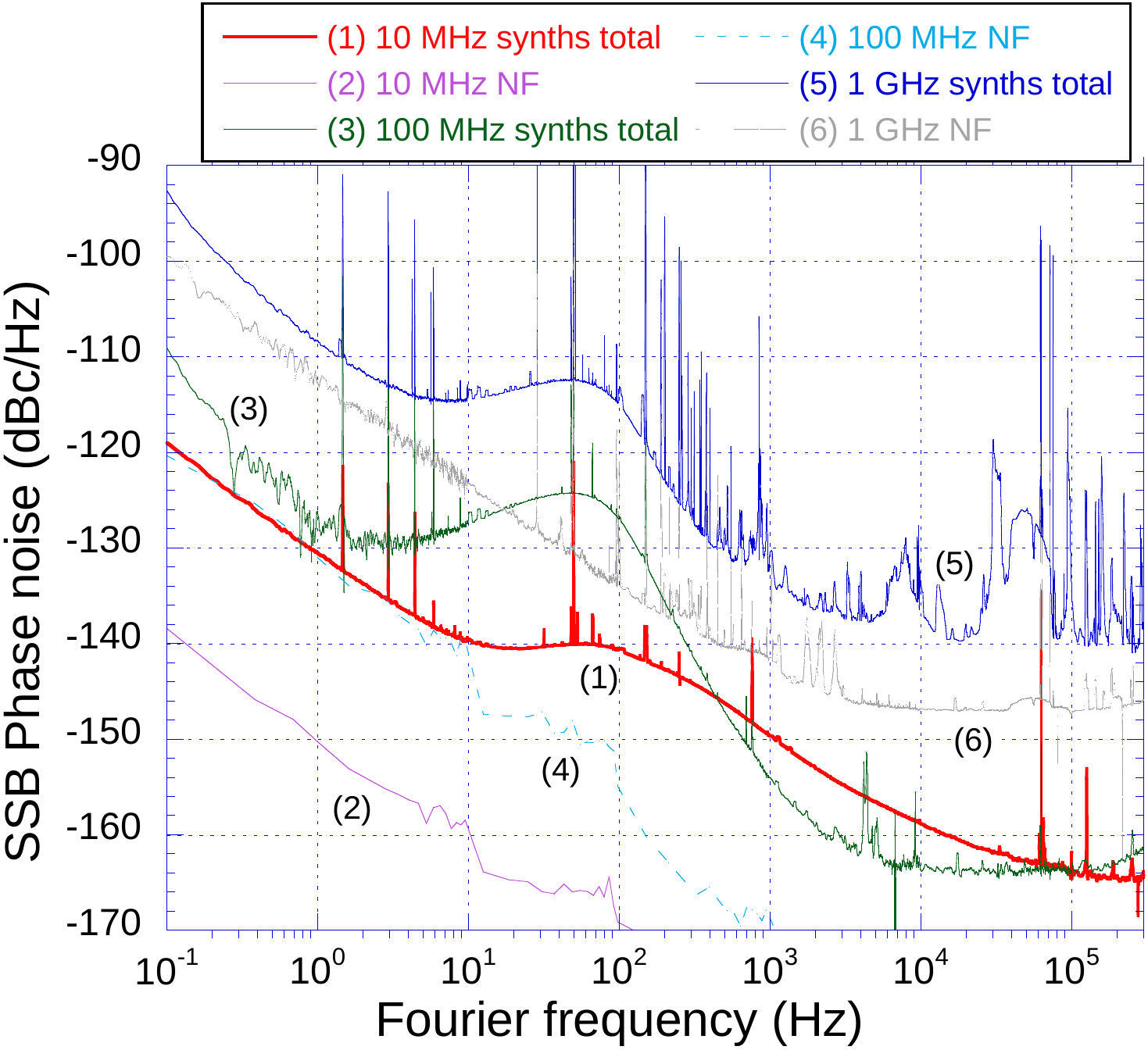}
\caption{Color online: Total SSB phase noise of the 10 MHz (curve 1), the 100 MHz (curve 3) and the 1 GHz (curve 5) signals synthesized from the 11.2 GHz output of an ultra-stable cryogenic sapphire oscillator. Each curve is shown as a comparison of two independent oscillator/synthesizers. For a single synthesizer 3 dB must be subtracted. Curves 2, 4, and 6 are the measurement system noise floors. }
\label{fig2}
\end{figure}

\begin{figure}[!t]
\centering
\includegraphics[width=3.5in]{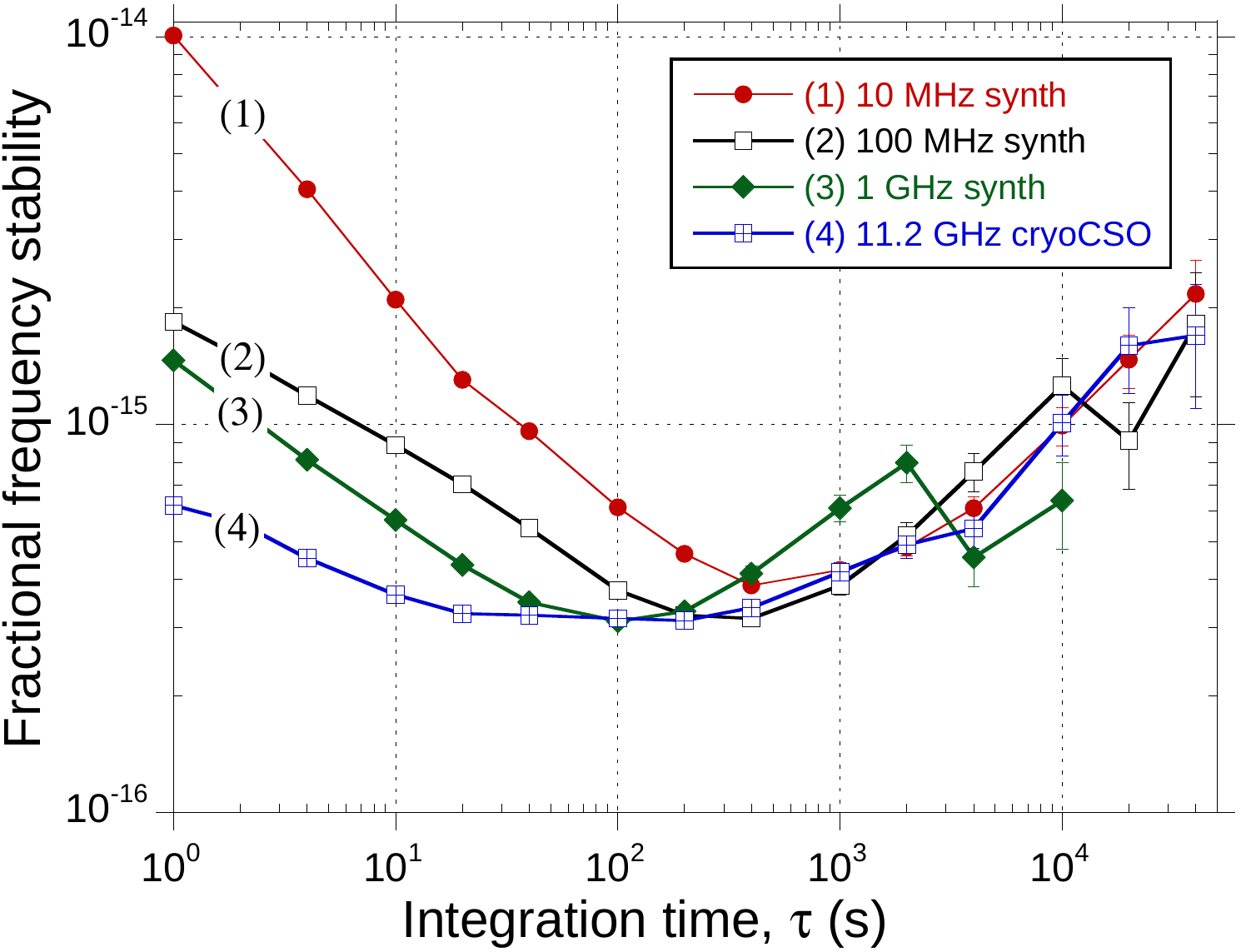}
\caption{Color online: The fractional frequency stability of 10 MHz (curve 1), 100 MHz (curve 2) and 1 GHz (curve 3) signals synthesized from the 11.2 GHz output of an ultra-stable cryocooled sapphire oscillator. Curve 4 represents the fractional frequency stability from a comparison between two independent 11.2 GHz cryocooled sapphire oscillators (cryoCSOs). Each curve represents the stability of a single oscillator/synthesizer but derived from a comparison of two nominally identical systems, where a factor of $1/\sqrt2$ has been applied assuming they are independent.}
\label{fig3}
\end{figure}

\subsection{Total phase noise of the 100 MHz and 10 MHz synthesizer with cryocooled oscillator}

The measurement of the total phase noise of the 100 MHz and 10 MHz signals was straight-forward and achieved by directly comparing the 100 MHz and 10 MHz output signals from the synthesizers (as indicated by the dashed-lined rectangle in Fig. \ref{fig1}) pumped with 11.2 GHz signals from independent cryocooled sapphire oscillators. Each comparison was made  using a Symmetricom 5125A test set with an input bandwidth of 400 MHz. Their resulting total phase noise and frequency stability are presented in Figs \ref{fig2} and \ref{fig3} respectively. Their measurement system noise floors were taken directly from the test set for the 100 MHz and 10 MHz measurements.

All measurements represented in Fig. \ref{fig3} are taken from the test set with a noise equivalent (NEQ) bandwidth of 0.5 Hz. Since the test set gives only a measurement noise floor, statistical error bars ($\pm \sigma_0$) have been estimated and applied to curves in the figure from the following. The number of ``samples'' for a given value of integration time $\tau$ is estimated by $N = 2 \, \tau_{max}/\tau$ where $\tau_{max}$ is the maximum value of $\tau$ generated by the test set for the given data set. The resulting errors are estimated by $\sigma_0=0.5 \times \sigma_y / \sqrt{N}$.

\subsection{Residual phase noise of the 1 GHz synthesizer components}

Using the baseband technique (as shown by the block diagram in Fig. 4(a)) we measured the residual phase noise contribution from the phase-locked 1 GHz Crystek VCO and the custom made PLL circuit by comparing the two similar systems. Fig. \ref{fig5} shows the Single Side Band (SSB) Power Spectral Density (PSD) of the relative phase fluctuations for two free-running Crystek VCOs (curve 1) drawn from a fit to our measurement data. This is compared with the measured residual phase fluctuations from 2 VCOs (curve 3) each phase-locked to the same 1 GHz signal from the frequency synthesizer. Curve 2 represents the modeled phase noise based on the free-running VCO phase noise and the measured transfer function of the PLL circuit with a gain bandwidth of 1 MHz.

The same technique was applied to two complete nominally identical 1 GHz synthesizers incorporated with the phase-locked VCOs as shown in Fig. \ref{fig1}.  As outlined in Fig. 4(b) two 1 GHz synthesizers were used to measure their relative residual phase fluctuations. The resulting residual phase noise was measured to be the same as curve 3 in Fig. \ref{fig5}; precisely laying on top, hence not shown for clarity. This is the same as just the residual phase noise of the VCO's phase-locked to the same 1 GHz source as outlined in Fig 4(a). This means that the residual phase noise contribution from the additional frequency synthesis components to generate 1 GHz from the 1.4 GHz and 100 MHz inputs adds negligibly to the residual phase noise of the phase-locked VCO.

\begin{figure}[!t]
\centering
\includegraphics[width=3.5 in]{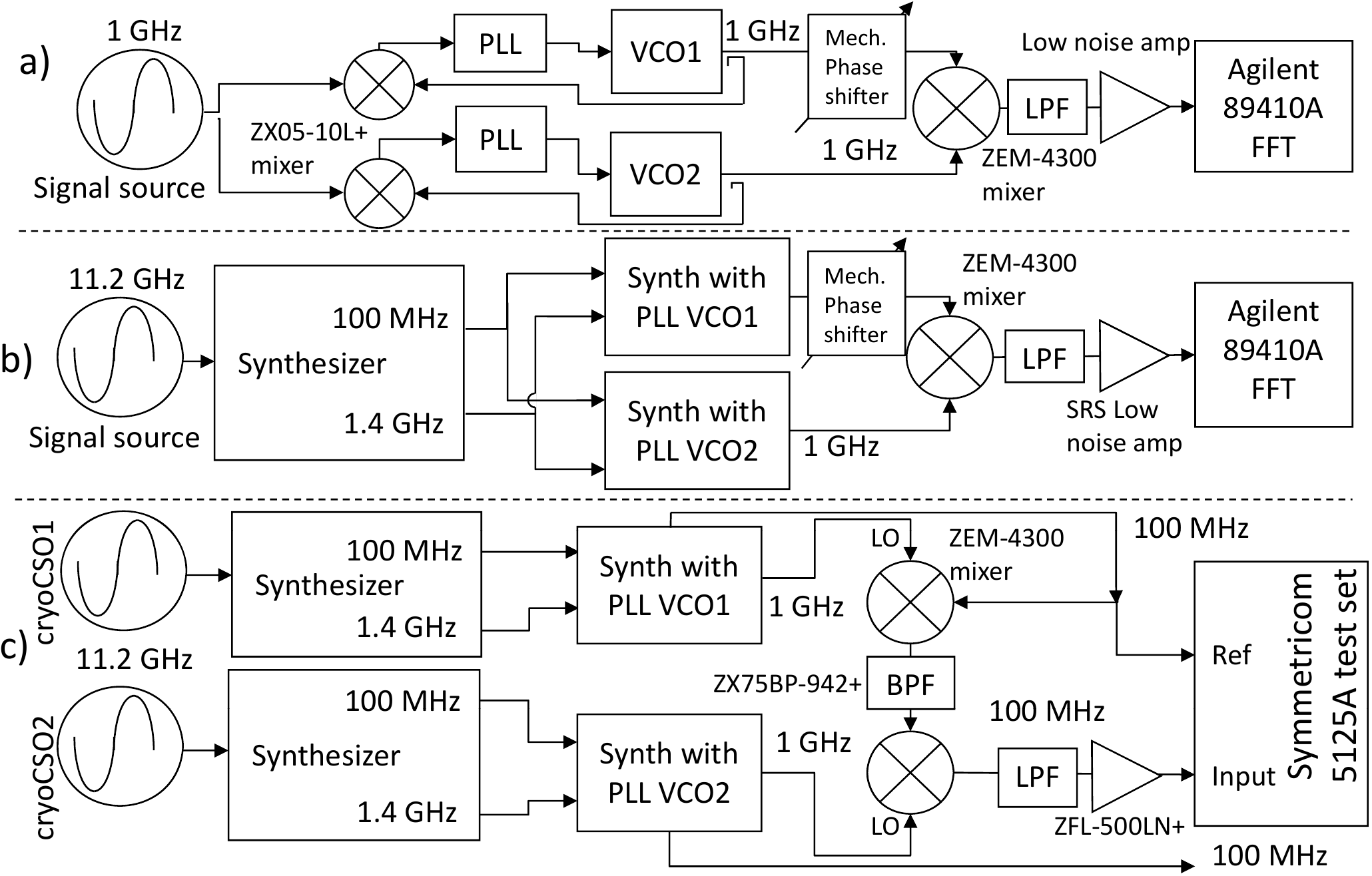}
\caption{Measurement methods: (a) Baseband technique used to measure the residual phase noise of two nominally identical phase-locked Crystek VCOs, (b) Baseband technique used to measure the residual phase noise of two nominally identical 1 GHz frequency synthesizers incorporating phase-locked Crystek VCOs and (c) New dual mixer method~\cite{Hartnett2013} used to measure the total phase noise and frequency stability of the 1 GHz synthesizers when supplied with an ultra-stable signal from independent cryocooled sapphire oscillators. LPF = Low Pass Filter. BPF = Band Pass Filter. MiniCircuits part numbers are indicated. }
\label{fig4}
\end{figure}

\begin{figure}[!t]
\centering
\includegraphics[width=3.5in]{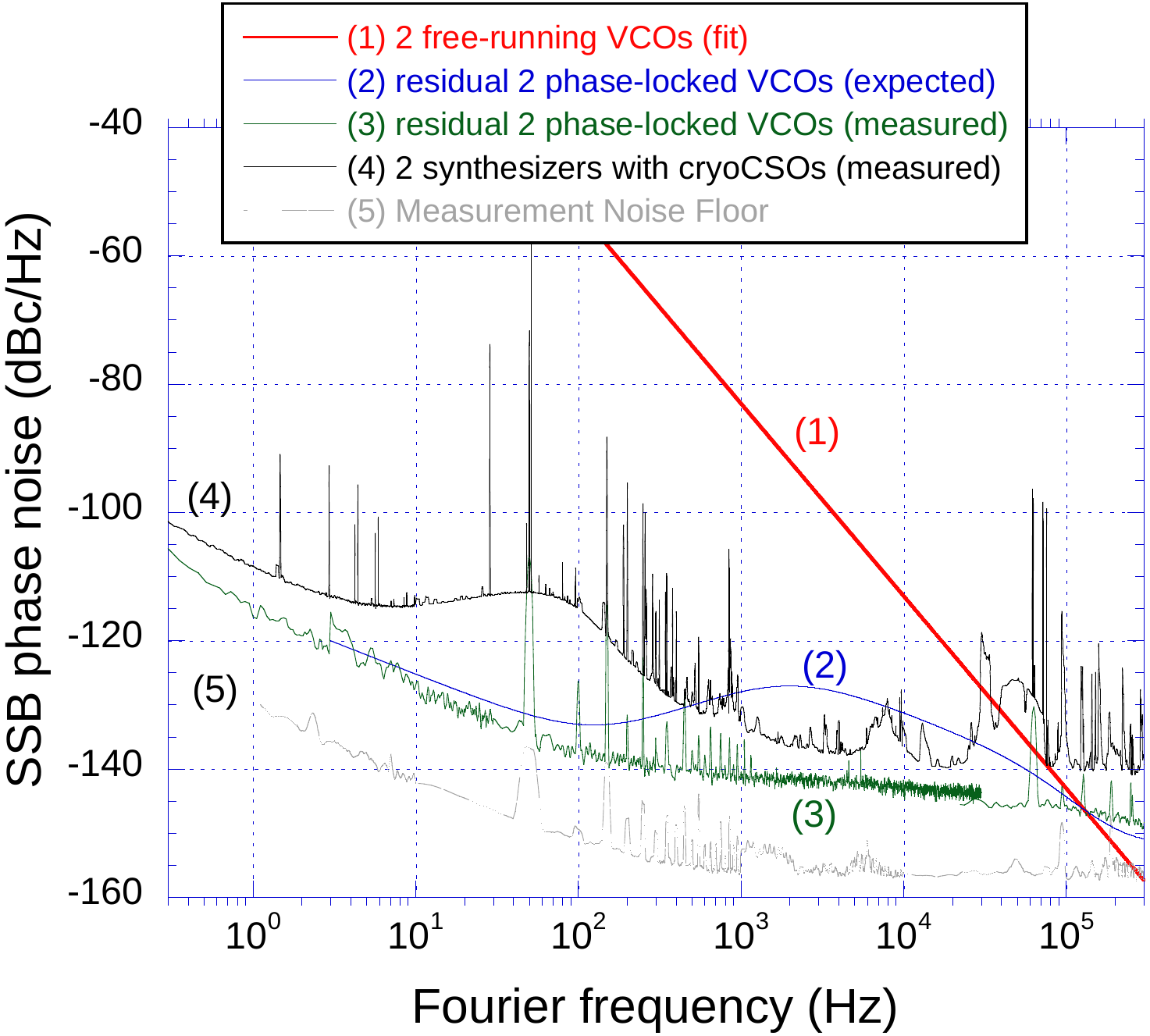}
\caption{Color online: PSD of phase fluctuations referred to a 1 GHz carrier. Curve 1 represents a fit of two nominally identical free-running Crystek 1 GHz VCOs. Curve 2 is the expected residual phase noise comparison of the VCOs phase-locked to a 1 GHz signal. Curve 3 is the measured residual phase noise comparison of the VCOs phase-locked to a 1 GHz signal. The residual phase noise for the latter plus the addition of the 1.4 GHz to 1 GHz synthesizer components exactly overlays curve 3, hence not shown for clarity. Curve 4 is the total phase noise for two nominally identical but independent systems comprising the cryocooled sapphire oscillator, the 11.2 GHz to 1.4 GHz and 100 MHz synthesizer, the 1.4 GHz to 1 GHz synthesizer and the 1 GHz PLL VCO. Curve 5 is the measurement system noise floor.
}
\label{fig5}
\end{figure}

\subsection{Total phase noise of the 1 GHz synthesizer with cryocooled oscillator}

In order to characterize the 1 GHz synthesizers driven by independent cryoCSOs a measurement technique is needed that allows one to measure their total phase fluctuations even though the cryoCSOs are subject to some frequency drift, albeit very small. We developed a method~\cite{Hartnett2013} that allows one to transfer the phase noise of the oscillators under test to a lower frequency signal that is within the bandwidth of a Symmetricom 5125A test set available in the lab.  The circuit diagram for the technique is outlined in Fig. 4(c), where a 900 MHz band-pass filter (BPF) and an auxiliary signal with a frequency ($f_a$) of 100 MHz, derived from the 100-MHz frequency synthesizer of one of the cryogenic sapphire oscillators, was used. See the full analysis and discussion in the Appendix. The resulting total phase noise and frequency stability of the 1 GHz synthesizer/oscillators are presented in Figs \ref{fig2} and \ref{fig3}, respectively.

The phase-noise-measurement noise floor was determined by disconnecting one oscillator/synthesizer from the measurement system (for example, the 1 GHz output of ``Synth with PLL VCO2'' in Fig. 4(c)) and the 1 GHz signal from the other synthesizer was power divided and the same signal sent as inputs to the LO ports of both mixers in Fig. 4(c). This then gives a noise floor that is largely the sum of the contributions from the mixers and the RF amplifier required to provide sufficient signal power to the test set. 

The spurs on curves 1, 3 and 5 in Fig. \ref{fig2} for offset frequencies $1 < f < 10$ Hz are due to the 1.4 Hz compressor cycle of the cryocooler. Most of the higher frequency spurs on curve 5 are also present in the measurement system noise floor and some arise due to imperfect filtering of the small switched power supply used in the 1 GHz VCO phase-locked synthesizer to bias the PLL circuit and the VCO. The hump near 70 kHz is from the Pound modulation frequency control system in the sapphire oscillators. 

The total phase fluctuations of the 1 GHz oscillator/synthesizers are compared with the residual noise measurements in Fig. \ref{fig5}, thus curve 5 of Fig. \ref{fig2} is reproduced in curve 4 of Fig. \ref{fig5}. At higher offset frequencies than shown in Fig. \ref{fig5}, outside of the locking bandwidth of the PLL control circuit, the phase noise is that of the free running Crystek VCO. According to the manufacturer's data this has a white noise floor near -170 dBc/Hz at 10 MHz. 

Lastly, using the method outlined in Fig. 2 of \cite{Hartnett2012} we measured the phase noise and frequency stability of the two independent cryocooled sapphire oscillators. Those results are, respectively, shown in curve 1 of Fig. 3 of \cite{Hartnett2012} and curve 4 of Fig. \ref{fig3} here.  These are the same oscillators used as the pump signals on the independent synthesis chains.

\subsection{Comparison of the 1 GHz oscillator/synthesizer with component phase noise contributions}

Referring to the synthesis chain shown in Fig. 4(c) the 11.2 GHz signal from the cryocooled sapphire oscillator~\cite{Hartnett2012} is combined with a DDS signal using a single side band mixer configuration, in the first synthesizer, to generate a signal frequency very close to 11.2 GHz with 14 significant figures~\cite{Nand2011}. This signal is subsequently used to generate both a 100 MHz and a 1.4 GHz signal that are in turn used to generate the 1 GHz signal in the second synthesizer. 

For each of these stages we have measured their phase noise. In the following we compare the measured total phase noise of the 1 GHz oscillator/synthesizer with the calculated total phase noise based on the sum of the measured phase noise in the components. 

The model we use is described by the following SSB power spectral densities (as functions of Fourier frequency, $f$) which sum to the total phase noise at 1 GHz according to,
\begin{eqnarray}
&\mathcal{S}^{T}_{1 GHz}(f) = \left(\frac{1.0}{11.2}\right)^2 \left [\mathcal{S}^{R}_{DDS}(f) +\mathcal{S}^{T}_{cryoCSO}(f) \right]  + \nonumber\\
&+4^2 \, \mathcal{S}^{R}_{100 MHz}(f)+ \mathcal{S}^{R}_{1.4 GHz}(f)+\mathcal{S}^{R}_{1 GHz}(f),
\label{eqn1}
\end{eqnarray}
where the superscript T and R represent total (or absolute) and residual (or additive) phase noise respectively. The subscript specifies the component contributing to the total 1 GHz oscillator/synthesizer phase noise. The factor of $(1.0/11.2)^2$ comes from the frequency division of the cryoCSO signal from 11.2 GHz  to 1 GHz. The factor of $4^2$ in front of the $\mathcal{S}^{R}_{100 MHz}$ term comes from the fact that the residual phase noise of the 100 MHz stage is frequency multiplied by 4 in the second synthesizer. See Fig 1. 


\begin{figure}[h]
\centering
\includegraphics[width=3.5in]{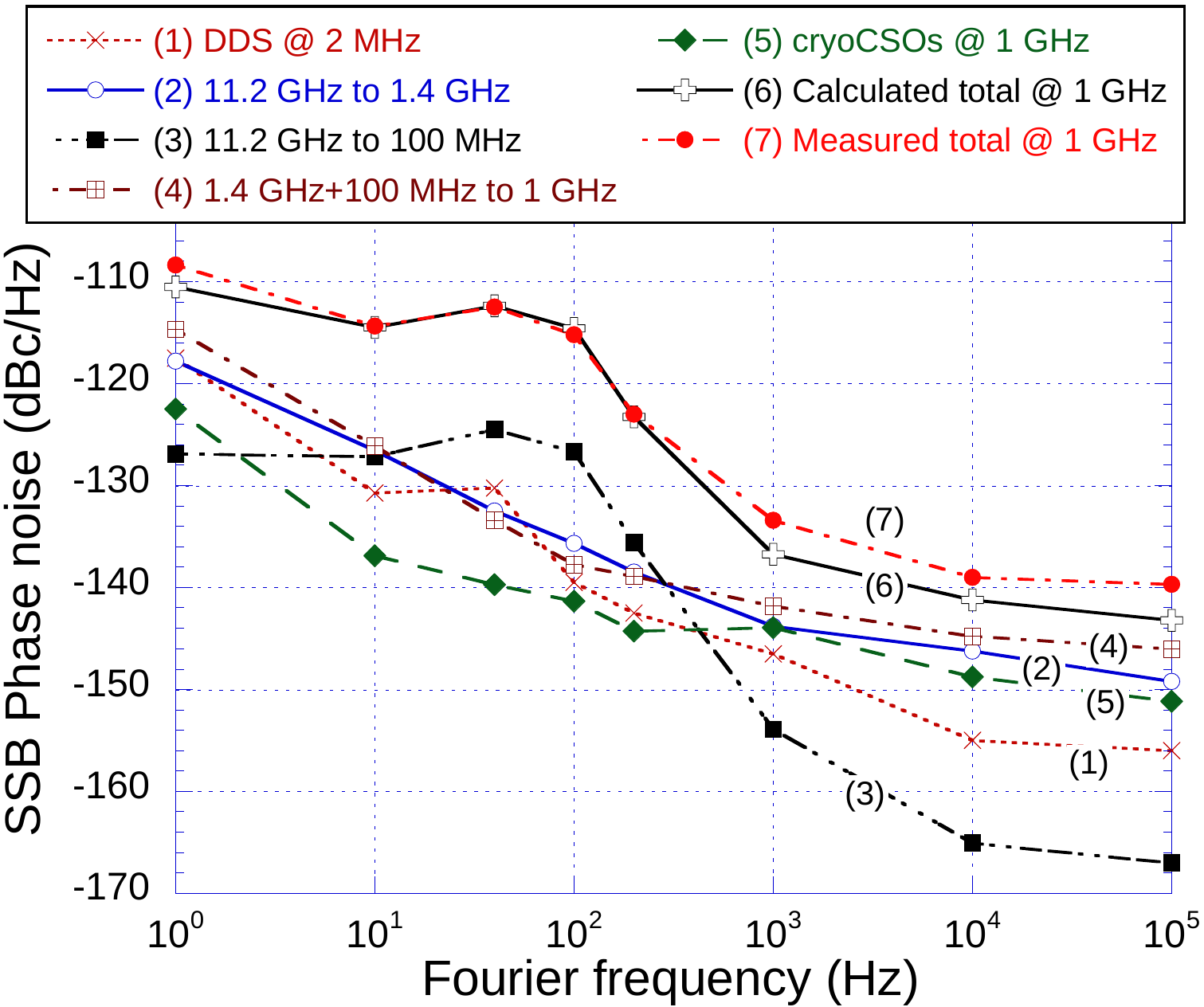}
\caption{Color online: A comparison of the measured phase noise of the different synthesizer stages involved with the production of the 1 GHz signal from the cryoCSO. Curves 2 and 3 are respectively the residual phase noise from the 11.2 GHz to 1.4 GHz divider stage and the 11.2 GHz to 100 MHz divider stage phase-locked to a 100 MHz quartz oscillator~\cite{Nand2011}. Curve 4 is the residual phase noise from the synthesizer described here in Fig. \ref{fig1}. Curve 5 represents the expected phase noise contribution from two independent cryoCSOs with the DDS contribution at 1 GHz, but scaled from 11.2 GHz~\cite{Hartnett2012} by subtracting approximately 20 dB from X-band phase noise data (from Fig. 3 of ~\cite{Hartnett2012}). Curve 6 is the calculated sum of the phase noise contributions from each of curves 2 to 5. Curve 7 is the measured phase noise of the 1 GHz synthesizer/oscillators from curve 4 of  Fig. \ref{fig5} (reproduced in curve 5 of Fig. \ref{fig2}). 
}
\label{fig6}
\end{figure}

The total phase noise of the 11.2 GHz cryoCSOs used in this calculation is shown in Fig. 3 of \cite{Hartnett2012}. The measured phase noise for each of the other 4 contributions are shown as curves 1 to 4 in Fig. \ref{fig6}, where each point represents the relative phase fluctuations of two devices. Data have been copied from the  measurement data ignoring any spurs. Table I lists these by curve number in Fig. \ref{fig6}, the component (oscillator or synthesizer), the type of phase noise, the method used to make the measurement (FFT means the standard baseband technique), the circuit used, and the source of the data used here. Curve 5 in Fig. \ref{fig6} is the calculated total phase noise contribution of the cryoCSO and the DDS unit to the 1 GHz oscillator derived from the first term in Eq. (\ref{eqn1}). Curve 6 is the calculated total phase noise of all contributions (from all terms in Eq. (\ref{eqn1})). The result agrees within a few dB of the measured data (curve 7). The agreement between curves 6 and 7 validates the new technique used here to measure oscillator phase noise. The new measurement technique was only used on the 1 GHz oscillator/synthesizer (curve 7). All contributions that sum to this (curve 6) were measured using previously proven techniques.

The increase in noise around 100 Hz is due to residual phase noise of the 100 MHz oscillator caused by insufficient locking bandwidth in the PLL used in the first synthesis stage~\cite{Nand2011}. It would be better to not use the quartz oscillator at all and this could be avoided. However one would need to give up the very low white phase noise floor seen on curve 3 in Fig. \ref{fig6} (or curve 3 of Fig. \ref{fig2} for the total phase noise), but this would have only a small impact on the white noise floor of the 1 GHz signal.

%
\begin{table}[!t]
\renewcommand{\arraystretch}{1.0}
\caption{Measured SSB Phase Noise Contributions to the 1 GHz Signal in Fig. \ref{fig6} and the Measurement Methods}
\label{table_1}
\centering
\begin{tabular}{|c|c|c|c|c|c|}
\hline
\#&Synthesizer/oscillator & PN & Method & Circuit & Data\\
\hline \hline
1&DDS @ 2 MHz   					& R  	& FFT  			& Fig. 1 in \cite{Nand2011}	 	& Fig. 3 in \cite{Nand2011}\\
2&11.2 GHz--1.4 GHz				& R 	& FFT 			& Fig. 1 in \cite{Nand2011}		&	Fig. 3 in \cite{Nand2011}\\
3&11.2 GHz--100 MHz 			& R 	& test set 	& Fig. 1 in \cite{Nand2011}		& Fig. 4 in \cite{Nand2011}\\
4&1.4GHz+100MHz--1GHz 		& R 	& FFT 			& Fig. 4(b)	here								& Fig. \ref{fig5} here\\
5&cryoCSOs @ 1GHz 				& T 	& test set 	& Fig. 2 in \cite{Hartnett2012}& Fig. 3 in \cite{Hartnett2012}\\
7&synth/cryoCSOs @ 1GHz 	& T 	& test set 	& Fig. 4(c)	here								& Fig. \ref{fig5} here\\

\hline
\end{tabular}
\end{table}

\subsection{Frequency stability of synthesizers compared}

Figure \ref{fig3} compares the fractional frequency stability (as Allan Deviation) of the 10 MHz (curve 1), 100 MHz (curve 2) and the 1 GHz (curve 3) signals synthesized from the cryocooled sapphire oscillator with that of the cryocooled sapphire oscillator itself (curve 4). A factor of $1/\sqrt 2$ has been applied to the measured data to represent the results for a single independent oscillator/synthesizer.  

The measured short term stability (integration times $\tau <100$ s) for the 10 MHz and 100 MHz signals is as expected from the residual phase noise of the frequency dividers measured in \cite{Nand2011}. However, the 1 GHz signal offers a small improvement over the 100 MHz signal, representing a factor of 1.6 better stability for $\tau = 10$ s.

The 1 GHz synthesizer is particularly sensitive to room temperature changes. The peak in the Allan deviation measurement near $\tau = 2 \times 10^3$ s is due to the air-conditioning cycle in the lab, manifesting as frequency fluctuations in the Crystek VCO itself. Initially the enclosure of the VCO was not thermally secured to the Aluminum baseplate in the box containing this synthesizer and the peak (in curve 3 of Fig. 5) was stronger and occurred near $\tau = 1 \times 10^3$ s. By securing the VCO to the baseplate the sensitivity was reduced and hence proves its origin. 

The following noise model was fitted to the data of curves 1 through 4 of Fig. \ref{fig3},
\begin{equation}
\sigma_y(\tau) = a_0 \tau^{-1}+ a_1 \tau^{-1/2} + a_2 + a_3 \tau^{1/2} +a_4 \tau, 
\label{noise model}
\end{equation}
with the best fit coefficients listed in Table II. Note where a dash appears in the table the coefficient was not used in the fit. 

The fit to the Allan deviation of the 10 MHz signal  was not particularly sensitive to the value of $a_2$. And distinctly it has a white phase noise term ($a_0$) in Eq. (\ref{noise model}) due to the white phase noise of the Holzworth 100 MHz to 10 MHz frequency divider used. A change in slope in curve 1 of Fig. \ref{fig5} is apparent at $\tau = 20$ s. 

Because of the peak in the Allan deviation data of the 1 GHz signal  near $\tau = 2 \times 10^3$ s  fitting to determine a linear frequency drift coefficient ($a_4$) was problematic, hence it can  only be estimated.

For integration times $\tau > 10^3$ s all signals are affected by the residual frequency drift between the two cryogenic sapphire oscillators as well as room temperature changes, including the air conditioning cycle with about a 20 minute period. Individually the cryogenic sapphire oscillators have an exponentially decreasing frequency drift and at the time the beat between the two cryoCSOs was measured their relative frequency drift was estimated from the above noise model (coefficient $a_4$ in Eq. (\ref{noise model})) to be $(1.4 \pm 0.3)\times 10^{-15}$/day, in addition to a random walk of frequency component. 

The random walk of frequency term ($a_3$) was included in the noise model fit to get best fits for the linear frequency drift term. Summing the latter with the former for all signals, with the exception of the 1 GHz signal, their Allan deviation evaluates to $\sigma_y \approx 4 \times 10^{-15}$ for $\tau = 10^5$ s -- a performance better than that of most hydrogen masers.


%
\begin{table}[!t]
\renewcommand{\arraystretch}{1.1}
\caption{Noise model Equation (\ref{noise model}) coefficients ($\times$ a multiplying factor) from fits to data of Fig. \ref{fig3}}
\label{table_2}
\centering
\begin{tabular}{|c|c|c|c|c|c|}
\hline
\#	&$a_0\, 10^{-15}$ & $a_1\, 10^{-15}$ & $a_2\, 10^{-16}$ & $a_3\, 10^{-18}$ & $a_4\,10^{-20}$\\
\hline \hline
1	&$4.9 \pm 0.3$	& $5.2 \pm 0.2$  	& - 							& $7.5 \pm 1.0$	 	& $1.7 \pm 0.6$\\
2	&-							& $1.7 \pm 0.06$  & $2.0 \pm 0.2$ 	& $2.7 \pm 0.7$	 	& $2.7 \pm 0.4$\\
3	&- 							& $1.3 \pm 0.02$ 	& $2.0 \pm 0.3$		& -								& $\sim 2$\\
4	&-							& $0.4 \pm 0.03$	& $2.3 \pm 0.2$ 	& $5.0 \pm 0.8$		& $1.6 \pm 0.3$\\

\hline
\end{tabular}
\end{table}

\section{Conclusion}

The design and evaluation of a 1 GHz synthesizer has been presented, wherein a low phase noise Crystek VCO is phase locked to a 1 GHz signal synthesized from an ultra-stable cryocooled sapphire oscillator. Given the high sensitivity of the VCO frequency  to temperature, future improvements in the long-term frequency stability of the synthesized signals should include direct temperature control of the VCO enclosure with a Peltier thermoelectric device and PID control. 

The phase noise and fractional frequency stability for the 1 GHz, the 100 MHz and the 10 MHz signals synthesized from two nominally identical cryocooled sapphire oscillators operating at 11.2 GHz have been presented. These results are compared and contrasted with those of the 11.2 GHz signal itself. The T4Science maser has a phase noise of -105 dBc/Hz on a 100 MHz carrier and -125 dBc/Hz on a 10 MHz carrier at 1 Hz offset~\cite{T4Science} compared to -130 dBc/Hz and -134 dBc/Hz, respectively, measured here for a single device.  

The frequency stability of the 100 MHz and the 10 MHz signals generated from the cryocooled sapphire oscillator, allowing for 1 month of aging, is superior to that from most hydrogen masers~\cite{T4maser}, even out to 1 day of averaging, where frequency drift has not been subtracted.

The 1 GHz signal offers nearly a factor of 2 improvement in the short term fractional frequency stability over that of the 100 MHz signal. This factor of 2 is close to the expected improvement given by the ratio of frequencies 1000 MHz/ (4 $\times$ 100 MHz). Any future improvement though could exclude the use of the 100 MHz quartz oscillator in the first synthesis stage. The  frequency dividers used there offer sufficiently low phase-noise performance at 100 MHz, and the elimination of the quartz oscillator would  reduce the complexity of the additional PLL.

Though the phase noise in the signals generated here is near state-of-the-art it is quite obvious that performance (as compared to the originating X-band signal) must be sacrificed through the synthesis process. This level of excess noise added by the frequency synthesis chain is clearly seen in the short term fractional frequency stability of the signals shown in Fig. \ref{fig3} and in the first two columns of Table II. Nevertheless, the performance of these signals  may open new horizons of research in frequency metrology, mmVLBI radio astronomy and  low noise radar where close-to-the-carrier phase noise is important.

\section*{Appendix}

\begin{figure}[!t]
\centering
\includegraphics[width=3.5in]{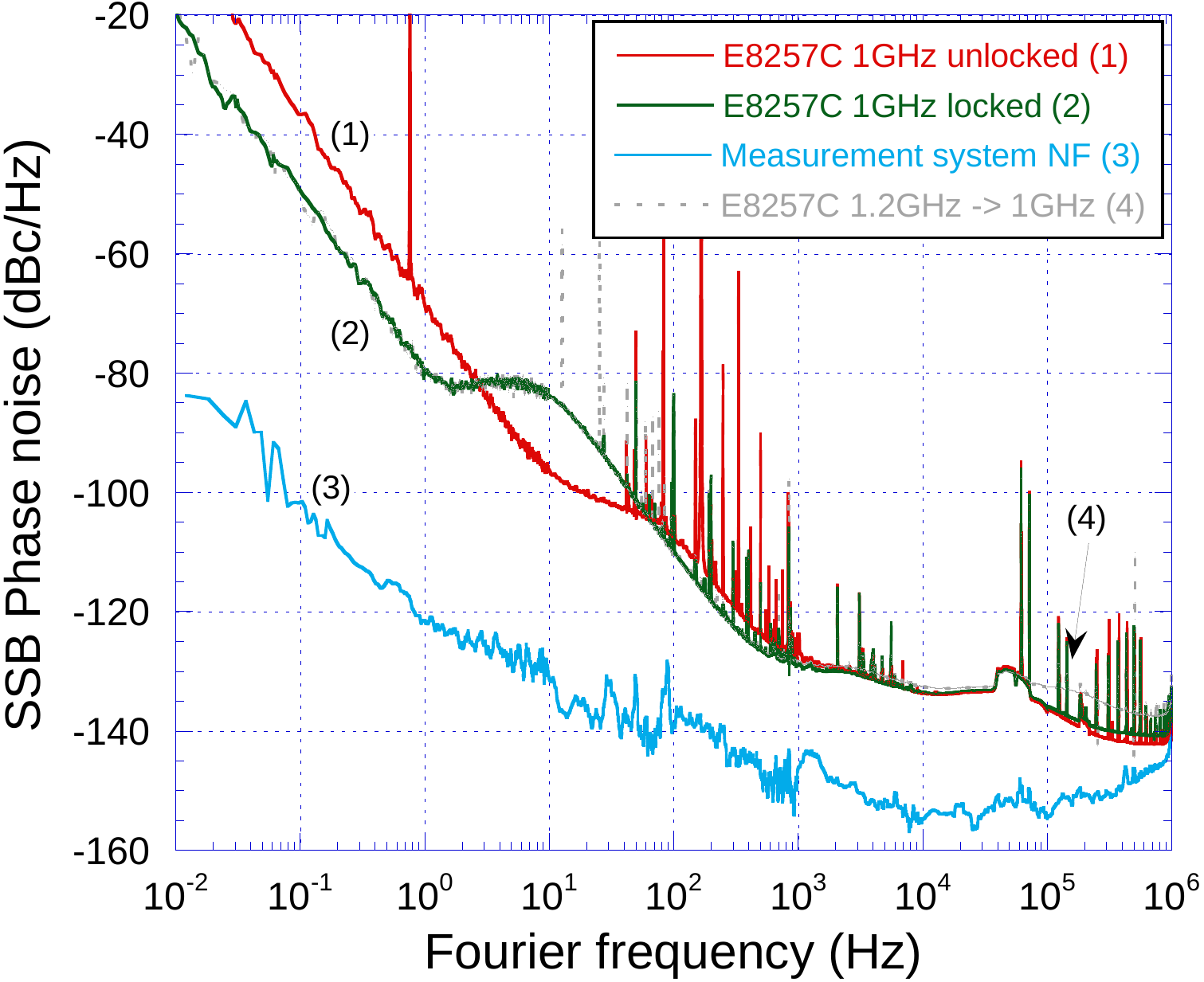}
\caption{Color online: The SSB phase noise of an Agilent E8257C synthesizer at 1 GHz without external reference (curve 1), and phase locked to a 10 MHz signal from the frequency doubled output of an Oscilloquartz 8607 quartz oscillator (curve 2). The latter was compared with a 1 GHz signal derived from an ultra-low phase noise cryogenic sapphire oscillator and measured using the measurement method of Fig. 4(c).  Curve 3 is the measurement system noise floor. Curve 4 is the phase noise of the  locked synthesizer at 1.2 GHz but referred to 1 GHz. See text for details.}
\label{figA1}
\end{figure}

\begin{figure}[!t]
\centering
\includegraphics[width=3.5in]{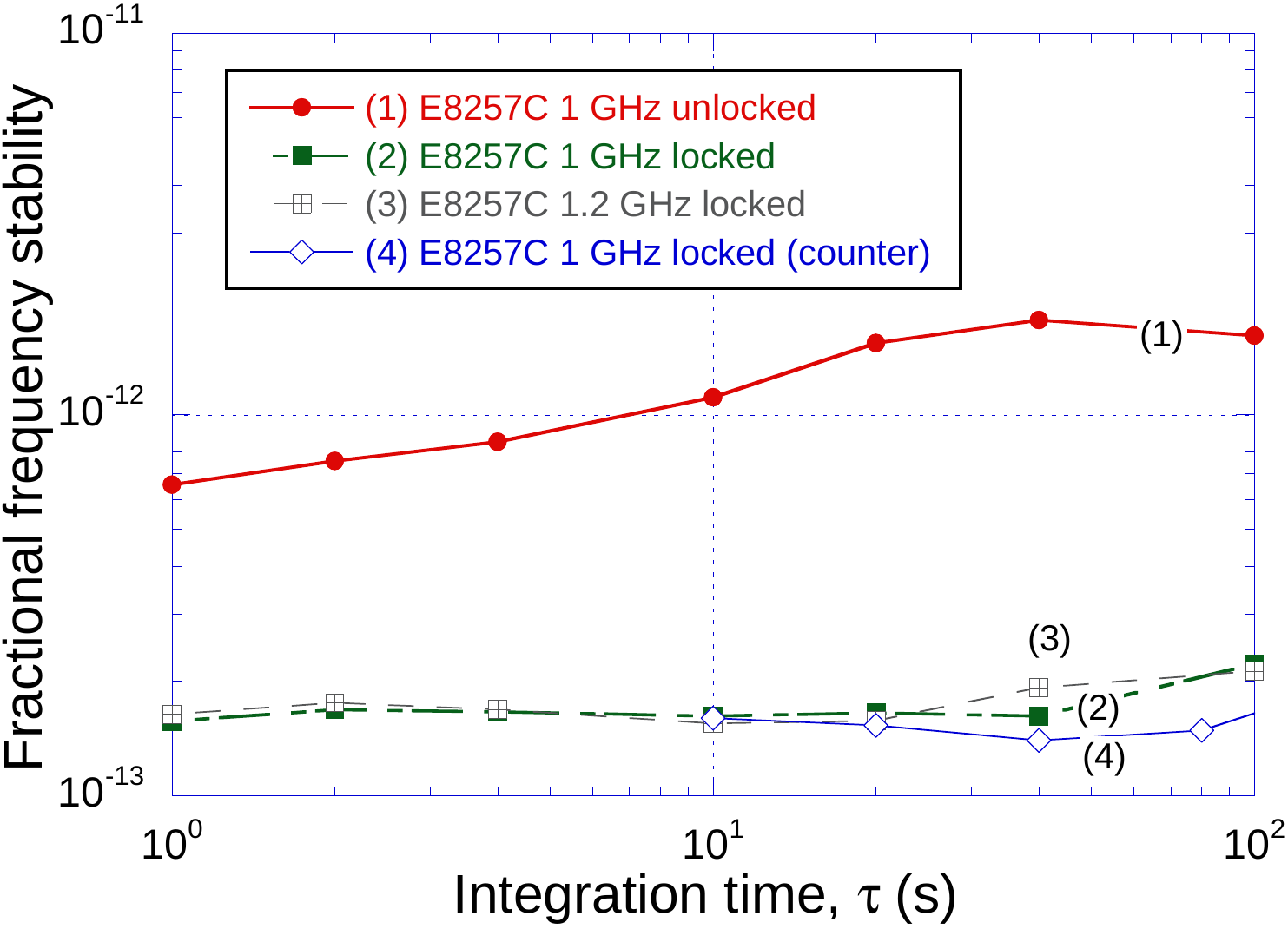}
\caption{Color online: Allan deviation of an Agilent E8257C synthesizer at 1 GHz without external reference (curve 1), and phase locked to a 10 MHz signal from the frequency doubled output of an Oscilloquartz 8607 quartz oscillator (curve 2). The latter was compared with a 1 GHz signal derived from an ultra-low phase noise cryogenic sapphire oscillator and measured using the measurement method of Fig. 4(c). Curve 3 is the Allan deviation of the signal frequency of the locked synthesizer at 1.2 GHz after the appropriate scaling was applied. See text for details. Curve 4 is the Allan deviation calculated from the beat of the locked synthesizer with a frequency of 1.006 GHz on a $\Lambda$-type frequency counter.}
\label{figA2}
\end{figure}

Figure 4(c) describes the basic features for a transposed frequency measurement method for the comparison of two oscillators operating at frequencies $f_{01}$ and $f_{02}$. We introduce a low frequency auxiliary signal $f_a$ and  a band-pass filter is used after the first mixing stage with the requirement that 
\begin{equation}
f_a \pm  |f_{01} - f_{02}| < 400 \,\, \verb1MHz, 1
\label{eqn400}
\end{equation}
with the $\pm$ depending on the choice of band-pass filter.  This ensures the generation of a signal at a frequency measurable by the test set containing the phase fluctuations of the two oscillators. Here we have assumed a 400 MHz bandwidth test set is being used, which is what we have available in the lab (a Symmetricom model 5125A). Therefore Eq. (\ref{eqn400}) means the frequency difference between the oscillators  must be less than twice the bandwidth of the test set otherwise additional mixing stages are needed.

In the case where one oscillator is a reference oscillator with  much lower phase noise than the oscillator under test the final result is that of the oscillator under test only. If the oscillators are nominally identical then we get the relative phase fluctuations of the two devices. This means 3 dB must be subtracted for the phase noise of a single oscillator.

From one of the oscillators an intermediate frequency is generated with an auxiliary oscillator in the first mixing stage, which is filtered to reject one of the sidebands with the band-pass filter, and then mixed in the second stage with the signal from the other  oscillator, low pass filtered, amplified and measured on the Symmetricom test set.   The final stage RF amplifier is needed to satisfy minimum input power requirements of the test set.

The phase noise is read directly from the test set, but to find the Allan deviation of the signal at frequency $f_{01}$ or $f_{02}$ one must scale by the ratio $(f_a \pm |f_{01}-f_{02}|)/f_{0}$ where $f_0$ is chosen as equal to either $f_{01}$ or $f_{02}$. 

We can model the results after the first mixing stage and band-pass filter as proportional to, 
\begin{equation}
cos[2 \pi (f_{01} - f_a)t + \delta \varphi_1 (t)+ \varphi_{LO1}],
\end{equation}
where upper $f_a$ sideband has been filtered out by the band-pass filter. The phase noise of the oscillator is represented by $\delta \varphi_1 (t)$. When this is mixed at the second mixing stage with the output of the second oscillator with its own phase noise $\delta\varphi_2 (t)$ at the  frequency $f_{02}$, and modeled as proportional to, 
\begin{equation}
cos[2 \pi f_{02} t +\delta \varphi_2 (t) + \varphi_{LO2}],
\end{equation}
the resulting signal frequency dependence can be modeled by the product of the  latter with the former as,
\begin{eqnarray}
&cos[2 \pi (f_{01} - f_a)t +\delta \varphi_1(t) + \varphi_{LO1}] \times \nonumber \\
&cos[2 \pi f_{02} t +\delta \varphi_2 (t) + \varphi_{LO2} ] = \nonumber \\
&\frac{1}{2}cos[2 \pi (f_{01}-f_{02}+f_a) t +\delta \varphi_1(t)-\delta \varphi_2 (t) + \Phi],
\label{eqn4}
\end{eqnarray}
where $\Phi = \varphi_{LO2}-\varphi_{LO1}$ is a phase constant and the high frequency  mixing product of order $f_{01}+f_{02}$ has been filtered out. The power of phase fluctuations of the resulting signal at frequency $f_{01}-f_{02}+f_a$ is equal to the  combined power of phase fluctuations of the individual microwave oscillators.

Of course the low pass filter must now pass the signal with frequency $f_{01}-f_{02}+f_a$, and this frequency must be within the bandwidth of the measurement test set. This then necessarily affects the frequency at which the phase noise is measured and the scaling ratio to calculate the Allan deviation becomes $(f_{01}-f_{02}+f_a)/f_{0}$. If the oscillators are nominally identical an additional $1/\sqrt2$ factor must be applied to get the stability of a single oscillator.

Using this measurement method (of Fig. 4(c)), as an independent test of the technique, we measured the phase noise of an Agilent E8257C synthesizer generating a 1 GHz signal compared to a 1 GHz signal synthesized from one of the ultra-low phase noise cryogenic sapphire oscillators. This means that in Fig. 4(c) cryoCSO2 and the synthesis chain was replaced with an Agilent E8257C synthesizer. The resulting phase noise spectra are shown in Fig. \ref{figA1}, and the frequency stability in Fig. \ref{figA2}. In Fig. \ref{figA1} curve 1 represents the phase noise of the synthesizer without external reference, and curve 2 when it was referenced by a 10 MHz signal from the frequency doubled output of an Oscilloquartz 8607 quartz oscillator~\cite{quartz10}. 

That measurement system used two MiniCircuits ZX05-10L-S+ mixers, a ZVBP-909-S+ band-pass filter, a SLP-100+ low pass filter and a ZFL-500LN+ amplifier. The measurement system phase noise floor is shown in curve 3 in Fig. \ref{figA1}. This was determined by using the same low phase noise cryocooled sapphire oscillator on both input ports (for cryoCSO 1 and 2 in Fig. 4(c)). The auxiliary oscillator frequency used in these measurements was $f_a = 100$ MHz, derived from one of the  cryocooled sapphire oscillators, with a phase noise of -130 dBc/Hz at 1 Hz offset~\cite{Nand2011, Hartnett2012}. In this case it is necessary to scale the resulting stability generated by the test set by $f_a/f_{0} = 100$ MHz$/1.0$ GHz $= 1/10$ where $f_{0} = f_{01}= f_{02} = 1$ GHz.

We also compared the same oscillators where we raised the output signal frequency of the  Agilent  E8257C synthesizer to $f_{02}=1.2$ GHz.  The auxiliary oscillator was still at $f_a = 100$ MHz, which was mixed with the signal at $f_{01} =1$ GHz from the ultra-low phase noise cryogenic sapphire oscillator, then we filtered out the upper sideband with the band-pass filter  leaving the lower sideband at 900 MHz to be mixed with the 1.2 GHz signal. After low-pass filtering (using a MiniCircuits SLP-400 filter) and amplification this resulted in a 300 MHz signal  that was measured by the test set. Curve 4 in Fig. \ref{figA1} is the result where a factor of $20 log(1.2)$ has been subtracted to compare the results all at 1 GHz. It is essentially identical with curve 2 as expected.  

And in Fig. \ref{figA2} we show the Allan deviation of the 1.2 GHz synthesizer signal where the correct scaling has been applied to the output of the test set.  In this case $(f_{01}-f_{02}+f_a)/f_{02} = 300$ MHz$/1.2$ GHz $= 1/4$. Only Allan deviation data with a NEQ bandwidth of 0.5 Hz are shown. Finally, we shifted the frequency of the locked Agilent synthesizer to 1.006 GHz and counted the 6 MHz beat (with a 10 s gate time) on an Agilent 53132A $\Lambda$-type frequency counter as an ultimate confirmation that the method works. The result is shown in curve 4 of Fig. \ref{figA2}.

The cryocooled sapphire oscillator's phase noise and frequency stability~\cite{Hartnett2012} are orders of magnitude lower than that of the Agilent E8257C synthesizer. At 1 GHz its phase noise is approximately equal to the measurement system noise floor (curve 3 or Fig. \ref{figA1}), hence it does not contribute to these results. Figs \ref{figA1} and \ref{figA2} therefore show only the performance of the  E8257C synthesizer. The conventional counter method produced the same expected stability as that from the new technique and both are equal to the known stability of our 8607 quartz oscillator.

It should be noted that when the frequency difference of the two oscillators falls within the bandwidth of the test set their phase noise can be measured by taking the beat note of the two signals and comparing it to a previously characterized low noise reference. To confirm the validity of our new transposed frequency technique we used this direct comparison method to repeat the measurements in Figs \ref{figA1} and \ref{figA2} and found no discrepancy in the results.

\section{Acknowledgments}
This work was supported by ARC grant LP110200142. The authors thank Sang Eon Park and Sang-Bum Lee from KRISS, Korea for their advice and assistance.

\end{document}